# Large thermoelectric efficiency of doped polythiophene junction: a density functional study

**Zahra Golsanamlou***, **Meysam Bagheri Tagani**, **Hamid Rahimpour Soleimani**

Address: Department of Physics, Computational Nanophysics Laboratory (CNL), University of Guilan, Po Box:41335-1914, Rasht, Iran.

Abstract- The thermoelectric properties of polythiophene (PT) coupled to the Au (111) electrodes are studied based on density functional theory combined with nonequilibrium Green function formalism. Specially, the effect of Li and Cl adsorbents on the thermoelectric efficiency of the PT junction is investigated in different concentrations of the dopants for two lengths of the PT. Results show that the presence of dopants can bring the structural changes in the oligomer and modify the arrangement of the molecular levels leading to the dramatic changes in the transmission spectra of the junction. Therefore, the large enhancement in thermopower and consequently figure of merit is obtained by dopants which makes the doped PT junction as a beneficial thermoelectric device.

## 1- Introduction:

$\pi$-conjugated molecules are one of the most interesting candidates for low cost electronics like photonic, spintronic and thermoelectric devices [1–7] because of their outstanding properties such as discrete energy levels, Coulomb correlations and interference effects .8–11 In addition, the properties of the molecules can be controlled through the doping [12–15]. Due to the extensive use of molecular devices based on p-conjugated molecules, the study of the thermoelectric properties of these devices has attracted many attentions [16–18]. The thermoelectric efficiency is the ability of the system to convert the generated thermal energy to useful electrical energy described by a dimensionless quantity known as figure of merit:

$$ZT = \frac{G_e S^2 T}{kT} \qquad (1)$$

where, $G_e, S, T$ and $\kappa$ are electrical conductance, thermopower or Seebeck coefficient, applied temperature and thermal conductance, respectively. The thermal conductance includes two parts: $\kappa_{el}$ and $\kappa_{ph}$ ($\kappa = \kappa_{el} + \kappa_{ph}$), where, $\kappa_{el}$ is the electron thermal conductance arises from the electronic part of the thermal transport and $\kappa_{ph}$ is the phonon thermal conductance obtained from the phononic part of the thermal transport. The aim of thermoelectric studies is to obtain the higher values for the ZT, so that the devices with ZT>1 are known as an efficient thermoelectric device.

Polythiophene (PT) is one of the famous and applicable p-conjugated molecules, with thiophene (T) rings are located in antiposition states to each other and has non-degenerated ground state. Because of its potential in integrating optical and electronic devices, it is important to investigate the thermoelectric properties of PT and its derivative based devices [19–22]. Asaduzzaman et al. studied the change in electronic properties of PT via replacing of some hydrogen atoms (H) by nitrogen atoms (N) and methyl ($CH_3$) group based on density functional theory (DFT) [23]. They found that the methyl-substitution has weaker effect on electronic properties of PT than nitrogen atoms. The electronic and structural properties of pure and doped PT in oligomer and periodic forms were investigated by Kaloni et al. [24] using DFT calculation. In their work, strong modification in electronic structure of PT was obtained by dopants. In addition, they studied the impact of dopant concentration which shows that the binding energy and HOMO-LUMO gap can alter with the number of the dopants.

In this study the PT is considered in coupling with two Au (111) electrodes. The dopants selected for the adsorption on PT oligomers are Li and Cl because of their less and more electronegativity, respectively and stable adsorbing on PT. The structural calculations are done on 1) undoped PT, 2) Li adsorbed PT and 3) Cl adsorbed PT, see Figure 1, whereas, the thermoelectric results are obtained for mentioned oligomers in presence of gold electrodes. The effect of molecular length and dopants concentration are also studied on the thermoelectric efficiency of the junction. To the best of our knowledge, there are no experimental and theoretical data for comparing our thermoelectric results, but there are some theoretical studies about the structure, energy levels and conductance of isolated PT in presence of dopants [24–28]. Therefore, we compare our results based on DFT calculations with them. In the following section of the paper, we present our computational details. The results are discussed in section 3 and finally our paper would be ended by short conclusion of this work.

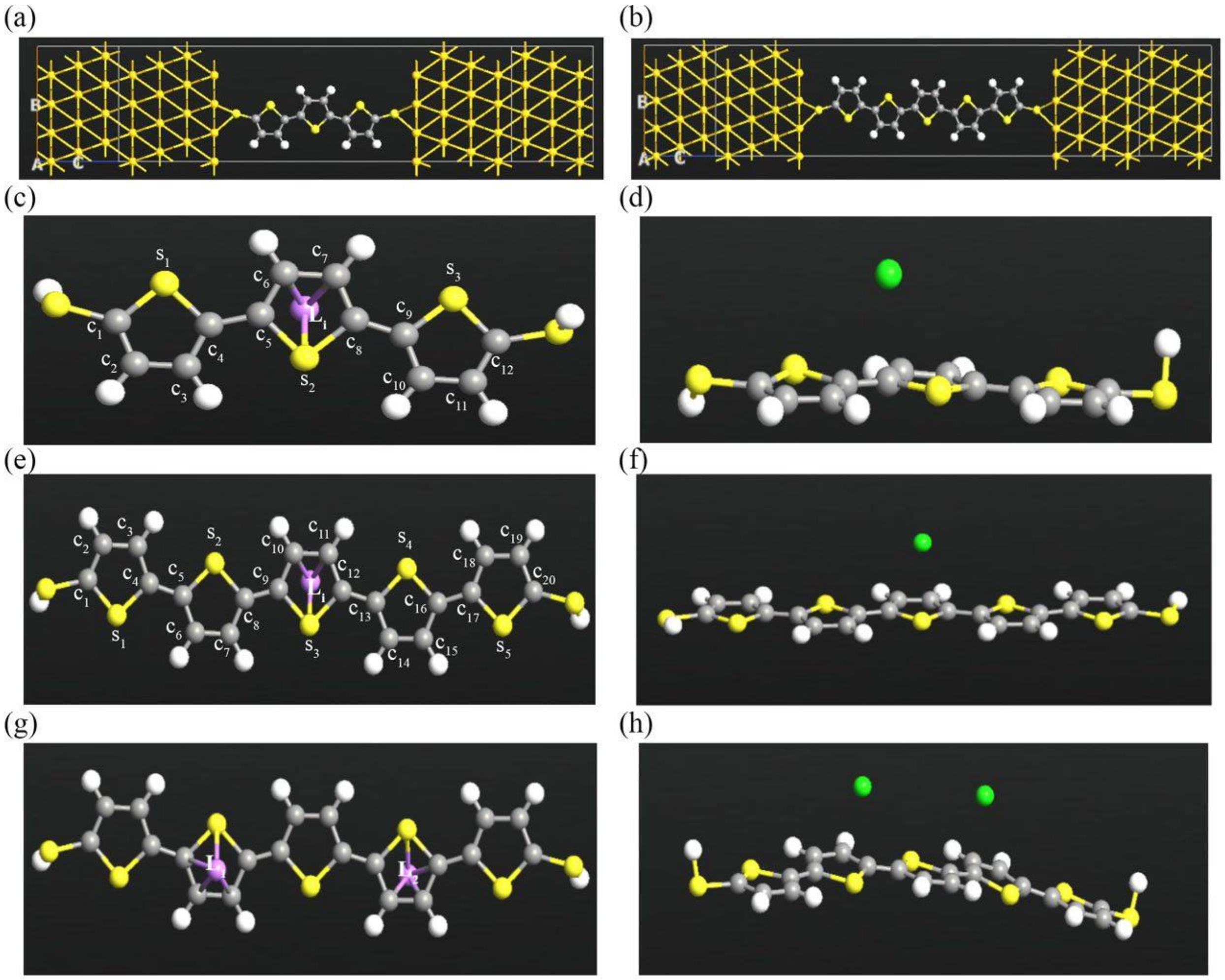

Figure 1: The polythiophene based molecular junctions: a) pure 3PT and b) pure 5PT. Doped polythiophene oligomers: c)1Li adsorbed on 3PT, d) 1Cl adsorbed on 3PT, e) 1Li adsorbed on 5PT, f) 1Cl adsorbed on 5PT, g) 2Li adsorbed on 5PT and h) 2Cl doped on 5PT. In the calculations, all oligomers are considered in coupling with Au (111) like parts a) and b). The Cl doped molecules are indicated in side view to show more clarity about their structure changes.

## 2- Computational methods

Our calculations, including structural relaxation and thermoelectric properties, are performed by SIESTA package based on DFT [29] within the generalized-gradient approximation (GGA) and parameterized correlation functional by Perdew-Burke-Ernzerhof (PBE) [30] We employ 2×2×100 k-point mesh, double zeta-single polarized basis set (DZP) for all atoms and 150 Hartree density cut off energy. The relaxation of the molecule with terminate groups, dopant atoms and molecule-electrode distance are performed until all forces converged to 0.02 eV/A° before coupling to the Au(111)-4×4 surfaces as a central region. The left and right periodic electrodes are attached to the central region of the junction and Au atoms are repeated by four times along the A and B directions. The length of the left and right electrodes is the same and equals to 7.0637365 A° . As thiol (SH) is the terminate group and the molecule tends to couple to the gold electrode in hollow position via sulfur atom [31,32], PT is connected to the gold electrodes in hollow position. The active region in our simulation is composed of the PT molecule and the four gold layers in each side. The transmission coefficient of the junction is obtained by TRANSIESTA code [33] to calculate the thermoelectric properties in linear response regime. In this regime, electronic and heat currents are expressed as follows [34]:

$$I = e^2 L_0 \Delta V + \frac{e}{T} L_1 \Delta T \tag{2a}$$

$$Q = e L_1 \Delta V + \frac{1}{T} L_2 \Delta T \tag{2b}$$

where, $\Delta T$ is the temperature gradient that can induce the voltage drop $\Delta V$. In above equations, $L_n$ are the coefficients obtained by the expansion of the electronic and heat currents [35] in terms of $\Delta T$ and $\Delta V$:

$$L_n = -\frac{1}{h} \int d\varepsilon (\varepsilon - \mu)^n \frac{\partial f}{\partial \varepsilon} T(\varepsilon) \tag{3}$$

where, $T$ and $\mu$ are the absolute temperature and chemical potential of the electrodes, respectively.

Also, $f(\varepsilon) = \left[1 + \exp\frac{\varepsilon-\mu}{kT}\right]^{-1}$ is the Fermi-Dirac distribution function of the electrode and $T(\varepsilon)$ is the transmission coefficient of the junction:

$$T(\varepsilon) = Tr\,[\Gamma_L(\varepsilon) G^r(\varepsilon) \Gamma_R(\varepsilon) G^a(\varepsilon)] \tag{4}$$

where, $\Gamma_L(\varepsilon)$ ($\Gamma_R(\varepsilon)$) describes the broadening function of the energy levels due to coupling of the left (right) electrode to the central region and $G^r(\varepsilon)$ ($G^a(\varepsilon)$) expresses the retarded (advanced) Green function of the junction. The thermoelectric coefficients are calculated as follows: the electrical and electron thermal conductances are [36–38]:

$$G_e = e^2 L_0 \tag{5a}$$

$$\kappa_{el} = \frac{1}{T}\left[L_2 - \frac{L_1^2}{L_0}\right] \tag{5b}$$

The thermopower is the proportion of the induced voltage drop to the applied temperature gradient when the current is zero:

$$S = -\frac{\Delta V}{\Delta T} = -\frac{1}{eT}\frac{L_1}{L_0} \tag{6}$$

where, e is the charge of an electron. Finally, the thermoelectric efficiency of the junction is obtained by Eq.1.

## 3- Results and Discussion

**A. Structural Results**: In this section we present the structural changes of the PT due to increase of the length and presence of the dopants. In going from 3T to 5T in pure PT, the S-C and C-C bonds inside the ring and C-C bond that connects two thiophene rings change about 0.001-0.01A. The dihedral angles have a bit change about 0.1-0.3. The next step is to consider the changes in the presence of the dopants. As it is observed from Figure 1c, e and g, the Li atoms are out of the thiophene rings, but in connection with three atoms and located at the center of the ring. Two Li atoms adsorbed on 5T are in opposite directions and thiophene rings are slightly twisted to pure ones. The outstanding changes in geometry of Li adsorbed PT are observed in bond lengths. The distances are in the range of 2.208-2.264 A for C-Li and 2.334-2.345 A for S-Li. When Cl atoms are adsorbed on PT oligomers, the structural changes are more significant.

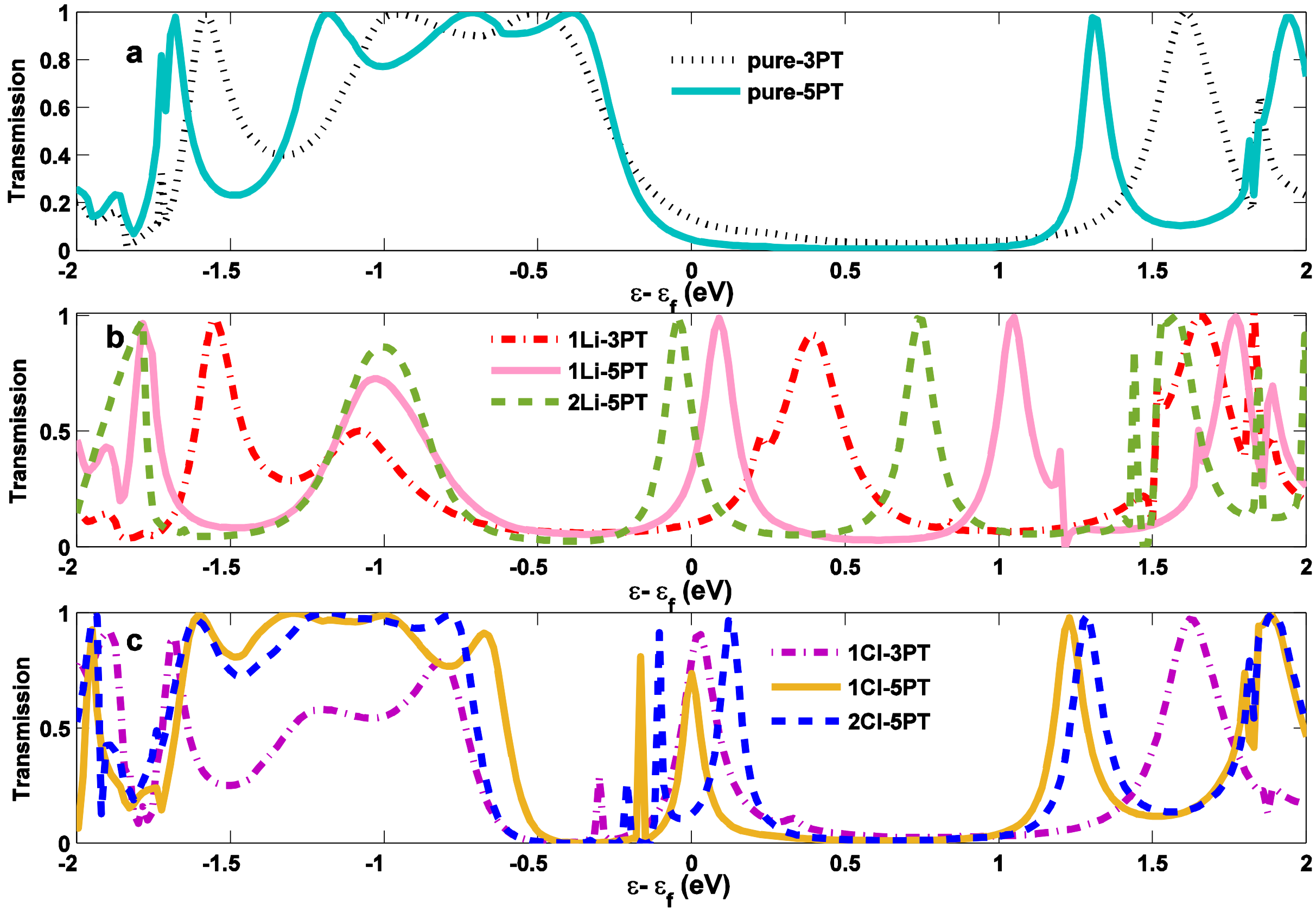

Figure 2:a) Transmission coefficient of considered junctions: a) pure PT junctions, b) Li-doped PT junctions and c) Cl-doped junctions. T=300K.

Unlike the Li adsorbed PT, the Cl atoms are not located above the center of the pentagon due to the repulsion between Cl and negatively charged C atoms around Cl atoms, which results in more asymmetry in bond lengths of Cl-doped thiophene rings. On the other hand, the Cl atoms are more electronegative than thiophene rings and get the negative charge from the pentagonal rings. The C-Cl distances for two nearest carbon atoms are found to be 2.78-3.28 A and S-Cl distances are between 3.858-3.946 A comparable with previous theoretical study [24] To make the analysis more accurate, we repeated the optimization of Cl-doped thiophene molecule with taking into account Van derWaals interactions using D3 Grimme approach [39]. Results show that the distances between Cl and adjacent carbon and sulfur atoms are slightly reduced, while the other distances remain the same as before. However, observed structural changes cannot induce a significant change in transport coefficient, so, in present work, we study the thermoelectric properties of the structures using GGA-PBE.

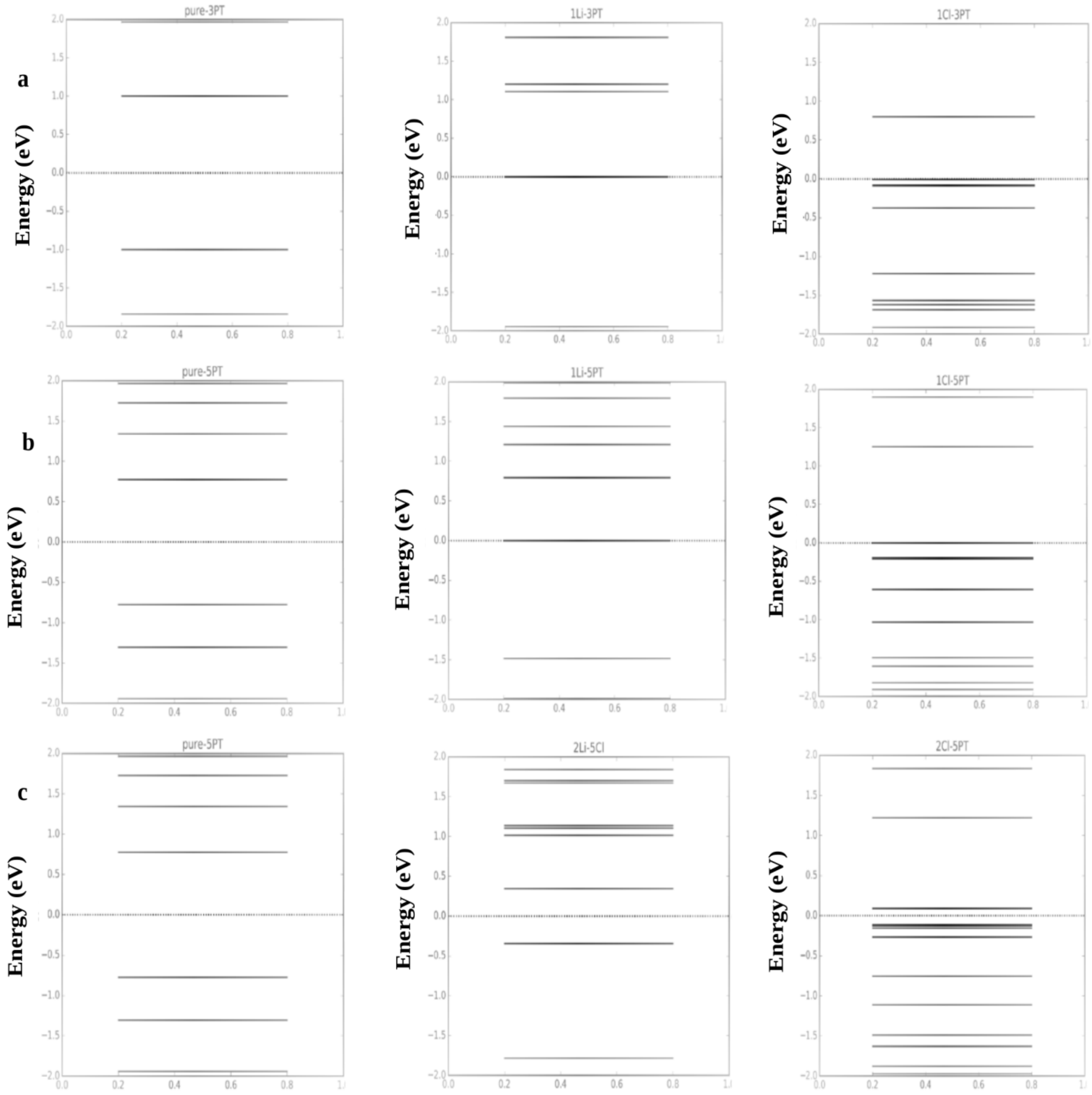

Figure 3 The molecular levels of isolated polythiophenes: a) pure and doped 3PT, b) pure and one adsorbed atom on 5PT and c) pure and two adsorbed atoms on 5PT.

By comparison of the electronegativity of the dopants and the thiophene rings, one can result that the molecule is n-doped in Li case and p-doped in Cl case. The Mulliken population analysis can help us to understand well the kind of doping. When Cl atoms are adsorbed on PT, 0.22e is transferred to Cl in 3T case and 0.37e in 1Cl-5T case, while the gained charge is different for the rings in 2Cl-doped 5T. The Cl atom located above the 2th ring obtains 0.38 electron, whereas the Cl located above 4th ring is charged about 0.56 electron [25]. The Cl atoms accept the charge from atoms in thiophene rings, so the atoms would have less charge than pure one and p-doping is formed. As the Li atom donates an electron to the thiophene ring, we have n-doping and the molecule is charged. Our analysis indicates that the Li atom loses 0.345 electron when 1Li atom is adsorbed on 3T. In addition, each Li atom donates 0.315 electron on 5PT and obviously, the atoms in the pentagon rings gain the donated charges. The fractional charge transfer is also observed in DFT studies of Li-doped structures [26,40].

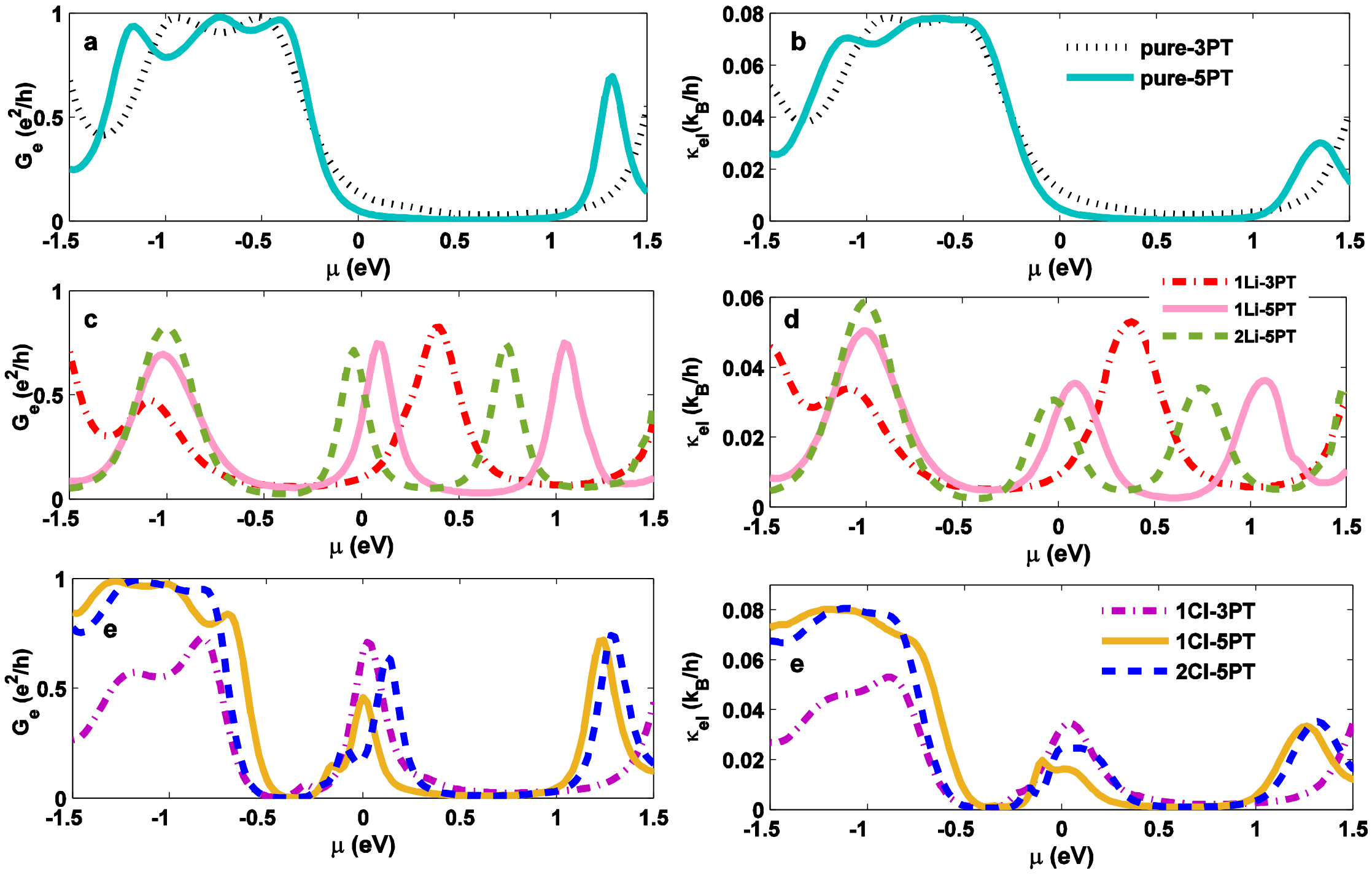

Figure 4: a) Electrical conductance of pure PT junctions, b) electron thermal conductance of pure PT junctions, c) electrical conductance of Li-doped PT junctions d) electron thermal conductance of Li-doped PT junctions, e) electrical conductance of Cl-doped PT junctions and f) electron thermal conductance of Cl-doped PT junctions, versus chemical potential of the electrodes. T=300K.

**B. Thermoelectric Results:** The transmission coefficient of pure and doped PT is indicated in Figure 2a-c. The molecular energy levels get closer when the molecular length increases. Therefore, the HOMO-LUMO gap decreases due to the electron donating feature of sulfur atoms and renormalization in molecular energy levels. It is also inferred that the HOMO-LUMO gap decreases in presence of the dopants with respect to the pure one [41,42]. When one of the HOMO or LUMO is located very close to the Fermi level ($\varepsilon - \varepsilon_f = 0$) the transport is in resonance limit. By contrast when the Fermi level is located between the HOMO-LUMO gap, the transport is in off-resonant limit (nonresonance tunneling). Regarding the Figure 2, PT in pure states is in off-resonant limit, but there is a resonance limit for tunneling in doped PT junctions. The change in molecular levels and consequently, tunneling regime alteration is the direct result of the modification of the electronic structures due to the doping [43–45]. Since the Li has the electron donating feature, the HOMO-LUMO gap shifts to the left side of the Fermi level in Li-adsorbed PT junctions and the LUMO level becomes closer to the Fermi level. While, the more electronegativity of the Cl causes the HOMO-LUMO gap shifts to the right side of the $\varepsilon - \varepsilon_f = 0$ and the HOMO level gets closer to the Fermi level ( Figure 2). For better comparison and more understanding, the HOMO, LUMO energies and the HOMO-LUMO gap of isolated molecules are indicated in the table 1. The broadening in energy levels in the Figure 2 is a result of the coupling of the molecules to the electrodes. When the molecule is doped, partially occupied molecular level is appeared in the electronic structure of the molecule leading to the HOMO gets closer to the Fermi level in the p-doped PTs and the LUMO gets closer to the Fermi level in the n-doped ones.

Figure 3 shows the molecular levels of isolated pure and doped PTs. It is clear that adding Cl or Li atoms opens new levels around the Fermi levels increasing the electrical conductance significantly.

In the case of Cl doped molecules, the more asymmetry is observed due to the significant changes in bond length and dihedral angles of the polythiophene causes to appearance some nearly degenerated energy levels. Spacing between these levels is less than broadening energy attributed to the coupling between the electrode and molecule. In these energy range, the electron wave can transfer through different levels, and as a result, constructive and destructive interferences may be happened. Constructive interferences lead to the maximum of the transmission, while destructive one reduces the transmission coefficient. On the other hand, such degeneracy is not in Li case, so, we expect that the transmission coefficient becomes maximum in the energy corresponding to the molecular energy levels. The predicted behavior is observed in transmission spectrum, Figure 2, and conductance in Figure 4 which is inferred as Fano resonances [46,47].

Table 1: The energies of HOMO, LUMO and the difference between them for isolated molecules

| **Molecule** | **HOMO** | **LUMO** | **HOMO-LUMO** |
|---|---|---|---|
| Pure 3PT | -1eV | 1eV | 2eV |
| Pure 5PT | -0.77eV | 0.77eV | 1.54eV |
| 1Li-3PT | -1.94eV | 0.00001eV | 1.94001 |
| 1Li-5PT | -1.48eV | 0.000013eV | 1.480013eV |
| 2Li-5PT | -0.344 eV | 0.342eV | 0.687eV |
| 1Cl-3PT | 0.00003eV | 1.5eV | 1.49997eV |
| 1Cl-5PT | -0.00005eV | 1.25eV | 1.25005eV |
| 2Cl-5PT | 0.09eV | 1.22eV | 1.13 eV |

Fermi energy is considered as reference energy and equals to zero.

The electrical and electron thermal conductances of polythiophene connected to the gold electrodes are indicated in Figure 4 versus chemical potential of the electrodes in both pure and doped PT. The both conductances decrease with increase of the molecular length owing to the reduction of the transmission at the Fermi energy [48]. It is also understood from the conductance plots of doped PT junctions that the charge transport increases in the Fermi energy and energies around it which is a gap in the pure PT junction. Because the electronic structure of the molecule is modified and the resonance peaks are appeared in $\varepsilon - \varepsilon_f = 0$ [49,50] so, the HOMO-LUMO gap decreases due to the decrease in the peaks distance. As the Fermi level is closer to the LUMO in Li-doped PT junctions, the electrons participate in the transport whereas the holes are responsible for the transport mechanism in Cl-doped ones due to the placement of the HOMO near the Fermi level. One of the important points is the presence of the antiresonance peak at the Fermi energy in the transmission coefficient of the 2Cl-adsorbed PT junction that causes minimum conductance in the Fermi energy of the junction. The conductance in Fano resonances is low because of the more asymmetry in the transport pathways (bond length) and dihedral angles of the doped molecules which results in destructive interferences in the resonance energies, so the conductance is decreased.

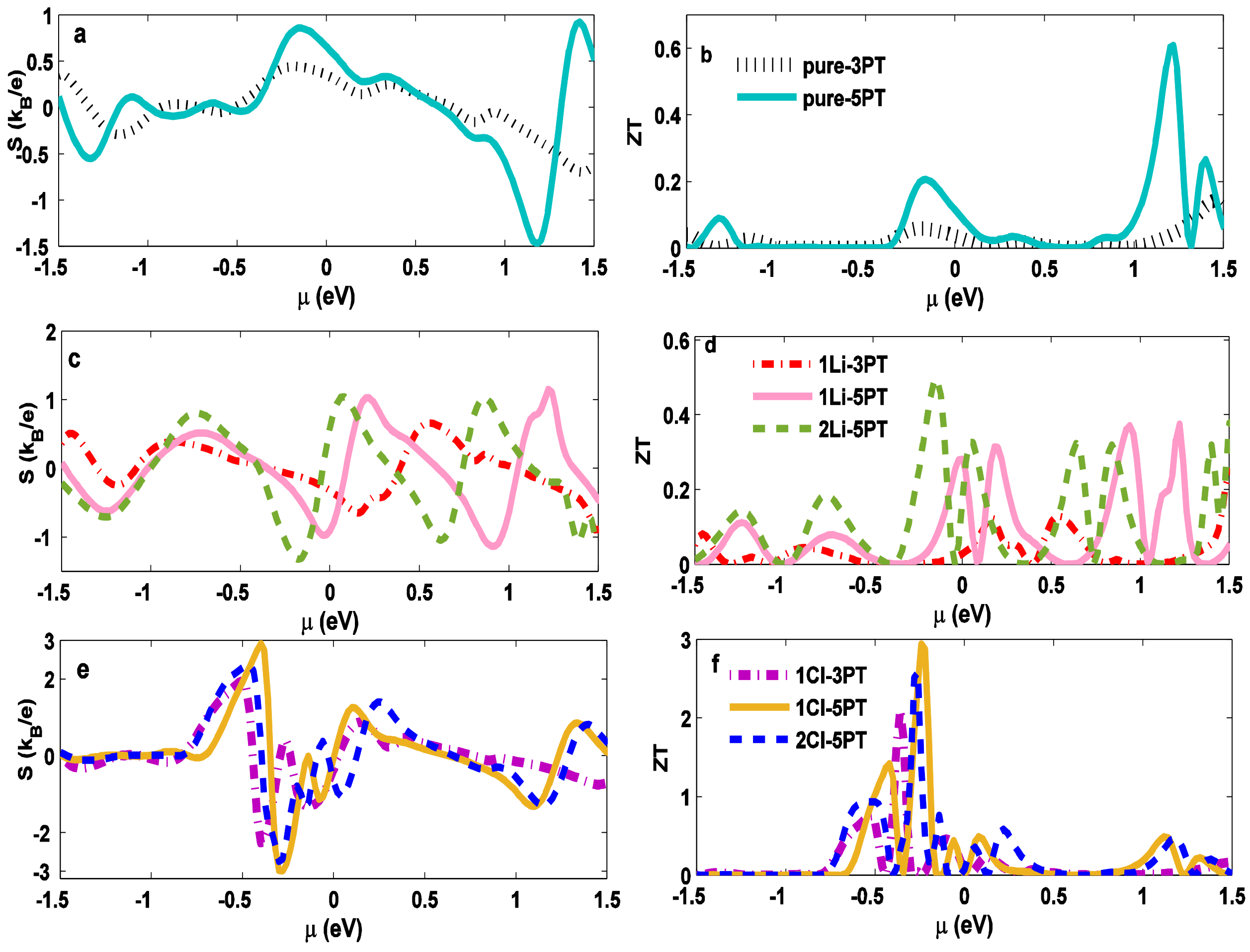

Figure 5: a) Thermopower of pure PT junctions, b) figure of merit of pure PT junctions, c) thermopower of Li-doped PT junctions d) figure of merit of Li-doped PT junctions, e) thermopower of Cl-doped PT junctions and f) figure of merit of Cl-doped PT junctions, versus chemical potential of the electrodes. T=300K.

Figure 5 a-h illustrates the thermopower (a, c and e) and figure of merit (b, d and f) of the considered junctions as a function of the chemical potential of the electrodes. The sign of the thermopower changes when the molecular levels cross from the Fermi level. The swings of the thermopower confirm the change in the electron population of the molecular levels [51,52]. These swings are more visible for the longer PT oligomers since by increasing the PT length, the number of p-electrons in the molecule increases. At low enough temperature, the Eq.6 can be simplified using Summerfield expansion about the Fermi energy [53,54]:

$$S = -\frac{\pi^2 k_B^2 T}{3e}\left(\frac{1}{T(\varepsilon)}\frac{\partial T(\varepsilon)}{\partial \varepsilon}\right)_{\varepsilon=\varepsilon_f} \tag{7}$$

By attention to the above equation, thermopower would have significant change in the energies whose transmission can be dramatically altered. Hence, the Fano resonances can produce both huge thermopower and figure of merit that result in high efficiency thermoelectric devices. Also, one can rewrite the Eq.6 to $S = -\frac{e}{T}\frac{L_1}{G_e}$ indicating the high thermopower for the energies with small $G_e$ [55–57]. Therefore, one can predict the large figure of merit near the points with low conductance. It is clearly observed from Figure 5e and f that the highest thermopower and figure of merit versus $\mu$ belong to the Cl-doped 5PT that have the Fano resonances near the Fermi energy.

In all considered junctions, the obtained value for the thermopower depends on the variation of the transmission coefficient. Figure 5c and e confirms that the presence of dopants increases the thermopower, as it is obtained in some experimental works and our previous work for the polyaniline junction [58,59]. In present study, it can be inferred from more change in the bond length and angles of the PT that causes more change in the slope of $T(\varepsilon)$. The positive/negative thermopower gives some information about the slope of transmission at the Fermi energy. The Li atoms have less effect on the structure of the PT due to their smaller size and lower electronegativity. So, the thermopower of the Li doped PT junction is less than Cl one but more than pure ones. Since the thermopower is present in ZT formula (Eq. 1) by order 2, it has more effect on the thermoelectric efficiency of the considered junctions. However, both electrical and thermal conductances should also be noticed.

At higher temperatures the role of the phonons in the thermal conductance is unavoidable [60,61]. At lower temperatures the contributions of the phonons can be neglected due to the big mismatch between the

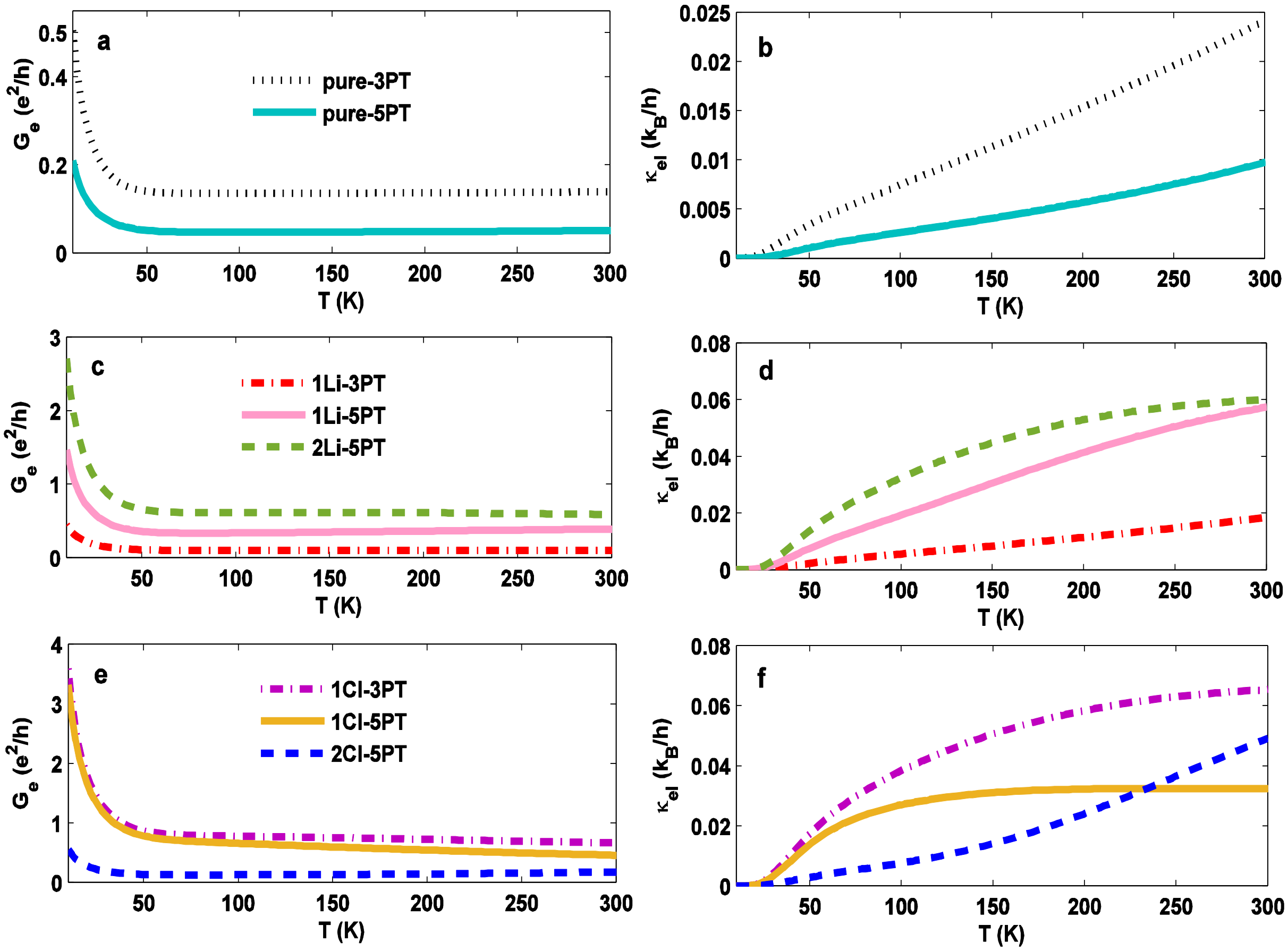

Figure 6: a) Temperature dependence of: a) electrical conductance of pure PT junctions, b) electron thermal conductance of pure PT junctions, c) electrical conductance of Li-doped PT junctions d) electron thermal conductance of Li-doped PT junctions, e) electrical conductance of Cl-doped PT junctions and f) electron thermal conductance of Cl-doped PT junctions.

phonon density of states of metallic electrodes and the phonon modes of the p-conjugated molecules at the metal-molecule interfaces resulting in low phonon transmission in the junction. The $ZT$ formula can be rewritten as $T = ZT_{el}\left(1+\frac{\kappa_{ph}}{\kappa_{el}}\right)^{-1}$ , where $ZT_{el}$ is the figure of merit of the pure electronic transport obtained by considering $\kappa_{ph} \approx 0$ [38,56,58,62,63]. In this study, we also ignore the phonon part of the thermal conductance and calculate the maximum value for the $ZT$. Figure 5b, d and f shows that the high thermoelectric efficiency is obtained where $k_{el}$ is low. It is understood from $ZT$ figures that the Cl-doped

PTs have the higher thermoelectric efficiency than other considered junctions and 1Cl-doped PT has the highest figure of merit near the Fermi level, where the Fano resonance is appeared. On the other hand, the Li-doped junctions have higher $ZT$ than pure junctions but less $ZT$ than Cl-doped ones. Therefore, one can anticipate that the presence of dopant improves the thermoelectric efficiency of the polythiophene based devices.

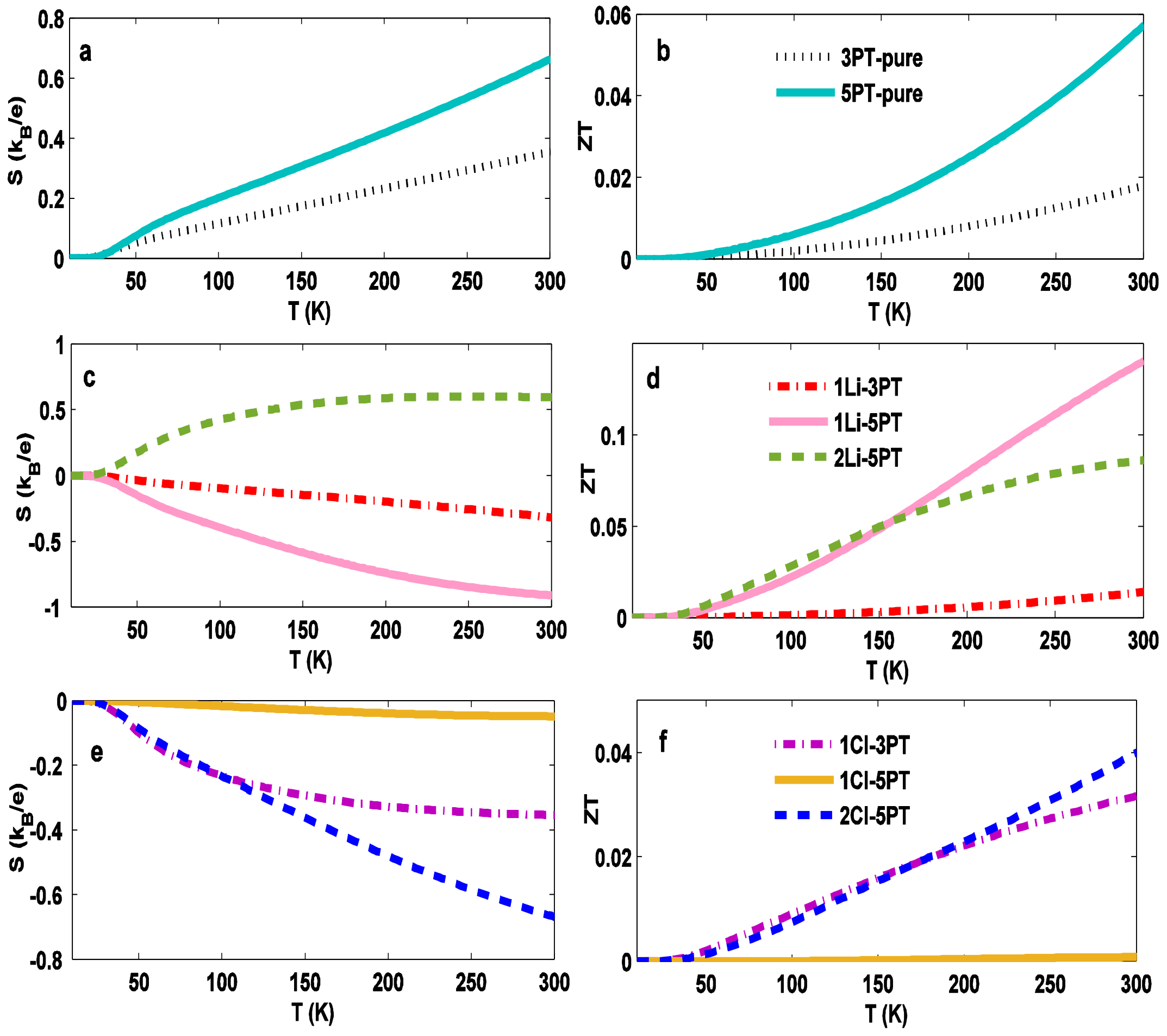

Figure 7: a) Thermopower of pure PT junctions, b) figure of merit of pure PT junctions, c) thermopower of Li-doped PT junctions d) figure of merit of Li-doped PT junctions, e) thermopower of Cl-doped PT junctions and f) figure of merit of Cl-doped PT junctions, versus temperature.

In the following, we study the temperature dependence of the thermoelectric coefficients of the considered junctions. Figure 6a, c and e illustrate the electrical conductance of the junctions versus temperature. As we mentioned in description of Figure 4, the electrical conductance decreases by the length enhancement. In the case of temperature, the electrical conductance decreases first and then increases by temperature. $T(\varepsilon)$ is independent of temperature but the Fermi derivative is a temperature dependent factor in $L_n$. By increase of temperature, the Fermi derivative is broadened and the height of it decreases. In higher temperatures, when the Fermi derivative $(-\frac{\partial f}{\partial \varepsilon})$ is broadened more part of $T(\varepsilon)$ can be entered in the thermal window $(-\frac{\partial f}{\partial \varepsilon})$, so, the decrease in height of $(-\frac{\partial f}{\partial \varepsilon})$ would be offset. The slight decrease of the $G_e$ in higher temperature for 1Cl-adsorbed PTs can be inferred from the symmetric transmission resonances with respect to the Fermi level that causes less value for the overlap of the T(ε) and the Fermi derivative in higher temperatures. As shown in Figure 6, the presence of dopants increases the conductance [64,65]. The main reason for this enhancement is the appearance of the new resonance and antiresonance peaks in $T(\varepsilon)$ due to the adsorb of dopants. Therefore, the distance of peaks decreases and the conductance can increase. It is worth noting that the increment in concentration of the dopants results in more enhancement in $G_e$ except in 2Cl-doped PT. 2Cl atoms cause more structural changes and as shown in Figure 1h, 2Cl-adsorbed 5PT is buckled which provides an internal electric field that is a barrier for the charge transport leading to the reduction of the conductance [66,67]. The electron thermal conductance of all studied junctions is shown in Figure 6b, d and f. Similar to the electrical conductance, the electron thermal conductance decreases by increasing the length. $\kappa_{el}$ is zero before 22K then increases by temperature enhancement. This increase is monotonic for the pure PT but in presence of dopants, $\kappa_{el}$ begins to rise from zero at 22K almost sharply and then reaches a uniform rise in high temperatures.

Figure 7 illustrates the thermopower and figure of merit of the studied structures versus temperature. The sign of thermopower in nonresonant limit can determine the kind of charge carrier participating in transport. The positive (negative) thermopower demonstrates the holes (electrons) responsibility for the transport mechanism whereas the sign of thermopower (+/-) in Figure 7 d-f can only point to the slope of transmission resonance obtained from the Summerfield expansion of S in Eq.7. Since the Fermi derivative broadens by the enhancement of temperature and can include more part of the transmission coefficient, the thermal induced transport rises and the thermopower increases consequently. In addition, the thermopower increases by the molecular length enhancement. A linear increase of the thermopower with molecular length was observed in previous theoretical and experimental studies [68–71]. The adsorb of dopants on PT has negative effect on temperature dependence of thermopower and reduces its magnitude. In presence of adsorbents, the distance of the energy levels from the Fermi level decreases but $L_0$ or $G_e$ increases (see Figure 6a, c and e). Therefore, the numerator becomes less than the denominator in Eq.6 and the thermopower decreases consequently. The lowest value for the thermopower appertains to 1Cl-5PT as a result of the compression of the peaks around the Fermi energy and less slope for the $\kappa_{el}$.

As it is mentioned above, we ignore the $\kappa_{ph}$ in this study because of the ignorable value of phonon transmission owing to the small value of the Debye frequency of the metallic electrodes with respect to the vibrational mode of the $\pi$-conjugated molecules. On the other hand, the obtained linear dependence of the thermopower on molecular length proves the existence of the little negative impact of the phonon transport on the thermoelectric efficiency. Hence, we report the maximum value of the figure of merit ($ZT_{el}$) as thermoelectric efficiency of the considered junctions in different temperatures. It can be observed from Figure 6 and Figure 7 that all thermoelectric coefficients are zero in temperatures less than 22K, because the width of $(-\frac{\partial f}{\partial \varepsilon})$ depends on temperature and in very low temperatures $(-\frac{\partial f}{\partial \varepsilon})$ is a delta function which cannot cover the T(e ). So, $L_n$ go to zero and thermoelectric coefficient becomes approximately zero. We can conclude from Figure 7b, d and f that the figure of merit versus temperature, increases by two factors: 1) increase of the molecular length and 2) adsorbing of the electron donating dopants. According to our analysis, the highest thermoelectric efficiency is obtained for 1Li-adsorbed 5PT, secondary 2Li-adsorbed on 5PT. Since the thermopower as a function of temperature is very low for 1Cl-doped 5PT, the lowest value for the $ZT$ belongs to this junction.

Among few experimental studies about thermoelectric properties of conducting polymers especially PT derivative materials, K. Hiraishi et al. have reported $ZT$ about $3 \times 10^{-4}$ for PT films at 293K [72]. J. Feng-Xing et al. have obtained $1.75 \times 10^{-3}$ for the figure of merit of PEDOT:PSS (Poly(3,4-ethylenedioxythiophene):Poly(styrenesulfonate)) at 270K [73]. Also, Yan et al. have obtained $ZT$ for acid doped polyaniline in the range of $0.7 \times 10^{-4}$ to $1.0 \times 10^{-3}$ at 313K [74]. On the other hand, some first principle studies have been investigated the thermoelectric efficiency of the organic materials. J. Chen et al. have demonstrated phthalocyanine crystals could have $ZT$ ranging from 0.2-2.5 at 298K [75]. Oligoyne molecular junctions at room temperature have illustrated $ZT = 1.4$ in the study of H. Sadeghi et al. [76]. In present study, the thermoelectric efficiency of polythiophene junction is significantly improved in presence of dopants specially Cl which $ZT$ rises to 3 in 1Cl-5PT junction at room temperature, see Figure 5, confirming a large efficiency for thermoelectric applications.

## 4- Conclusion

In this study, we have investigated the thermoelectric properties of doped polythiophene (PT) coupled to Au (111) electrodes based on density functional theory using nonequilibrium Green function formalism in linear response regime. We have considered Li atoms as a n-doping and Cl atoms as a p-doping in different concentrations for two lengths of the PT. The results indicate the that the molecular levels and the HOMO-LUMO gap of the molecule are significantly modified using the mentioned adatoms. Also, the structural changes like the change in bond length and dihedral angles result in the dramatic changes in the tunneling pathways so that the Fano resonances are observed in the transmission spectrum. We can observe the large value for thermopower and thermoelectric efficiency of the doped junction specially in the Fano resonances that provides a useful view to consider a doped PT junction as a good thermoelectric device.